 \documentstyle[prl,aps,twocolumn,floats,epsf]{revtex}

\newcommand{\beq}{\begin{equation}} 
\newcommand{\eeq}{\end{equation}} 
\newcommand{\beqa}{\begin{eqnarray}} 
\newcommand{\eeqa}{\end{eqnarray}} 
\newcommand{\mx}{\left[\begin{array}} 
\newcommand{\finmx}{\end{array}\right]} 
\newcommand{\mxp}{\left(\begin{array}} 
\newcommand{\finmxp}{\end{array}\right)} 
\newcommand{\casos}{\left\{\begin{array}} 
\newcommand{\fincasos}{\end{array}\right.} 
\newcommand{\rcasos}{\left.\begin{array}} 
\newcommand{\rfincasos}{\end{array}\right\}} 

\def\lsim{\ \rlap{\raise 3pt \hbox{$<$}}{\lower 3pt \hbox{$\sim$}}\ }
\def\gsim{\ \rlap{\raise 3pt \hbox{$>$}}{\lower 3pt \hbox{$\sim$}}\ }
\def\vev#1{\langle #1 \rangle}

\def\nnu{\nonumber}

\def\ea{{\it et. al.}}

\begin{document}

 \draft

 \title{Non anomalous $U(1)_H$  gauge model of flavor } 
\author{Jesus M. Mira, \ Enrico Nardi} 
\address{Departamento de F\'\i sica,  
Universidad de Antioquia, A.A.{\it 1226}, Medell\'\i n, Colombia}
\author{and \\ Diego A. Restrepo}
\address{Departament de F\'\i sica Te\`orica
IFIC-CSIC, Universitat de Val\`encia, Burjassot, Val\`encia {\it 
46100},Spain} 


%
\maketitle
\begin{abstract} 
A non anomalous horizontal $U(1)_H$ gauge symmetry can 
be responsible for the fermion mass hierarchies of the 
minimal supersymmetric standard model. Imposing the 
consistency conditions for the absence of gauge 
anomalies yields the following results:  
{\it i)} 
unification of  leptons and  down-type quarks 
Yukawa couplings  is allowed at most 
for two generations.
{\it ii)} 
The $\mu$ term is necessarily somewhat below the
supersymmetry breaking scale.
{\it iii)} 
The determinant of the quark mass matrix vanishes,  
and there is no strong $CP$ problem. 
{\it iv)}  
The superpotential has accidental $B$ and $L$ 
symmetries.
The prediction $m_{\rm up}=0$ 
allows for an unambiguous 
test of the model at low energy.
\end{abstract} 

 \medskip



One of the most successful ideas in 
modern particle physics is that of local 
gauge symmetries. A huge amount of data 
is beautifully explained in terms of the 
standard model (SM) gauge  group   
$G_{SM}=SU(3)_C\times SU(2)_L \times U(1)_Y$.
Identifying this symmetry required 
a lot of experimental and theoretical efforts, 
since $SU(2)_L \times U(1)_Y$ is  hidden 
and color is confined.
Today we understand  particle interactions 
but we do not have any deep clue in understanding 
other elementary particle properties, like 
fermion masses and mixing angles.
The SM can only  accommodate but not explain 
these data. 
Another puzzle is why $CP$ is preserved by 
strong interactions to an accuracy  $<10^{-9}$.  
One solution is to postulate that 
one quark is massless, but within the SM  
there are no good justifications for this. 
Adding supersymmetry  does not 
provide us with any better understanding 
of these issues. In contrast, it adds new problems.  
A bilinear coupling for the down-type and up-type
Higgs superfields $\mu\phi_d\phi_u$ is allowed both 
by supersymmetry and by the gauge symmetry. However,  
phenomenology requires that $\mu$ should be 
close to the scale where these symmetries 
are broken. With supersymmetry, several operators 
that violate baryon ($B$) and lepton ($L$) 
numbers can appear. However, none of the effects expected 
from these operators has ever been observed. 
Since a few of them can induce fast proton decay, 
they must be very suppressed or absent. 

Relying on the gauge principle, in this paper 
we attempt to gain insight into these problems.  
We extend minimally $G_{SM}$ with a 
{\it non anomalous} horizontal Abelian 
$U(1)_H$ factor. 
An unambiguous prediction of the non anomalous 
$U(1)_H$ is  a massless up-quark. 
This represents {\it the} crucial 
low energy test of our framework.
Shall future lattice computations rule out
$m_{\rm up}=0$ \cite{Cohen:1999kk}, the whole idea
will have to be abandoned.
 
To explain the fermion mass pattern we follow the 
approach originally  suggested by Froggatt and Nielsen 
(FG) \cite{Froggatt:1979nt}.  
$U(1)_H$ forbids most of the fermion 
Yukawa couplings. The symmetry is spontaneously broken 
by the vacuum expectation value (VEV) of a SM singlet 
field $S$, giving rise to a set of effective operators 
that couple the SM fermions to the electroweak Higgs field.  
The hierarchy of fermion masses results from 
the dimensional hierarchy among the various higher order 
operators. This idea was recently reconsidered  
by several groups, both in the context of supersymmetry 
\cite{Leurer:1993wg} and with a gauged $U(1)_H$ 
\cite{Ibanez:1994ig,Binetruy:1995ru,GaugeU1,Nir:1995bu}. 
It was argued that consistency  with phenomenology  
implies that $U(1)_H$ must be anomalous, and  
thus only the anomalous case was studied in detail. 

Our theoretical framework is defined 
by the following assumptions: 
\ 1.  
Supersymmetry and the gauge group 
$G_{SM}\times U(1)_H$.   
\ 2.  
$U(1)_H$ is broken only by 
the VEV of a field $S$ with horizontal 
charge $-1$\footnote{
We assume that a tree level Fayet-Iliopoulus $D$-term 
triggers the breaking of $U(1)_H$ 
while preseving supersymmetry.}.
$S$ is a SM singlet and 
is chiral under $U(1)_H$.  
\ 3. 
The ratio between the VEV  $\vev{S}$ and the 
mass scale $M$ of the FN fields is of the order of 
the Cabibbo angle  $\lambda \simeq \vev{S}/M \sim 0.2$. 
\ 4. 
The only fields chiral under $U(1)_H$
and charged under $G_{SM}$ are the 
minimal supersymmetric SM supermultiplets.  
\ 5. 
The lepton and down-type quark mass matrices  
$M^\ell$ and $M^d$ satisfy   
$\det M^\ell \leq \det M^d$  
(of course  this last assumption is an experimental fact).   

In the following we will use the same 
symbol to denote a field and its horizontal charge. 
Upon $U(1)_H$ breaking, the Yukawa 
couplings $Y^u\,$, $Y^d$ and $Y^\ell $ of the up-type  and 
down-type quarks and of the leptons are generated.  
They satisfy the following relations:
\begin{eqnarray}
\label{yukawa}
Y^u_{ij}&=&\ 
\casos{ll}
A^u_{ij}\> \lambda^{Q_i+ u_j+\phi_u}
& 
\ {\rm if } \quad 
Q_i+u_j+\phi_u\geq0,\\
0&
\ {\rm if } \quad 
Q_i+ u_j+\phi_u<0,
\fincasos
\end{eqnarray}
and similar ones for $Y^d$ and $Y^\ell$. 
The zero entries arise from  holomorphy, while   
$A^u_{ij}$ are numerical coefficients 
of order $\lambda^0$ that we will often leave
understood. Let us introduce the following 
combinations of charges: 
%
\beqa
\label{nf}
\begin{array}{lll}
n_u=\sum_i (Q_i+ u_i), 
& n_d=\sum_i (Q_i+ d_i),  
& n_Q=\sum_i Q_i, 
\phantom{\Bigg| }   \\
n_\ell=\sum_i (L_i+ \ell_i), 
& n_\phi=\phi_u+\phi_d, 
& n_L=\sum_iL_i. 
\end{array}
\eeqa
After electroweak symmetry breaking, the Yukawa
couplings (\ref{yukawa}) give rise to the fermion mass 
matrices $M^u$, $M^d$ and $M^\ell$. 
In the absence of  vanishing 
eigenvalues their determinants read 
\begin{eqnarray}\label{detm}
\label{detmu}
\det M^u&=&\ \langle\phi_u\rangle^3
\lambda^{n_u+
3\phi_u}\>\det A^u , \\
\label{detmd}
\det M^d&=&\ \langle\phi_d\rangle^3
\lambda^{n_d+3\phi_d}\>\det A^d ,
\\
\label{detml}
\det M^\ell&=&\ \langle\phi_d\rangle^3 
\lambda^{n_\ell+3\phi_d}\>\det A^\ell . 
\end{eqnarray}
Since all the entries in $A^{u,d,\ell}$
are of order $\lambda^0$,  $\det A^{u,d,\ell}$ 
is of order 1. Then the size   
of the determinants (\ref{detmu})-(\ref{detml})
is fixed by the horizontal charges and by 
the ratio of the Higgs doublets VEVs 
$\tan\beta =\vev{\phi_u}/\vev{\phi_d}$.

The SM Yukawa operators are invariant  
under a set of global $U(1)$ symmetries: 
$B$, $L$, hypercharge ($Y$) and a  symmetry 
$X$ with charges $X(d)=X(\ell)=-X(\phi_d)$
and $X=0$ for all the other fields.  
Therefore, shifts of the horizontal charges  
proportional to $L$, $B$, $Y$ and $X$ do not 
affect the fermion mass matrices.
In the following, we will denote as {\it equivalent}    
two sets of  charges  that 
can be transformed one into the other by means 
of shifts of this kind.
Note that the superpotential term 
$\mu\, \phi_u\phi_d$ (the $\mu$-term) 
is not invariant under $X$, and hence it can be 
different for two equivalent sets. 
Experimental evidences  
for non-vanishing neutrino mixings 
\cite{Fukuda:1998mi}  imply  that 
shifts proportional to individual lepton 
flavor numbers $L_a$ ($a=e,\mu,\tau$) 
transform between  phenomenologically 
{\it non equivalent} set of charges. In fact,    
while these shifts do not affect the charged 
lepton masses, they still produce 
different patterns of neutrino mixings.
In our analysis we will work with the 
following linear combinations of generators: 
$X$, $B$, $B$-$L$, $L_\tau$-$L_\mu$, 
$L_\mu$-$L_e$, and $Y$. 

Since $G_{SM}\times U(1)_H$ is a local symmetry,  
it is mandatory to study the (field theory) consistency 
conditions for cancellation of the gauge anomalies. 
The mixed $SU(n)^2\times U(1)_H$ anomalies,  
quadratic in $SU(n)=SU(3)_C\,,SU(2)_L\,,U(1)_Y$  
and linear in the horizontal charges,  
can be expressed in terms of the coefficients 
\beqa
\label{coefficients}
C_3&=&n_u+n_d, \nnu \\
C_2&=&n_\phi+(3n_Q+n_L),\\ 
C_1&=&n_\phi+ \hbox{$\frac83$}n_u+
\hbox{$\frac23$}n_d+2n_\ell -(3n_Q+n_L)\,. \nnu
\eeqa
The coefficient of the mixed 
$U(1)_Y\times U(1)_H^2$ anomaly  
quadratic in the horizontal charges reads 
\beq
\label{quadratic} 
C^{(2)}= \phi_u^2 - \phi_d^2 + 
\sum_i \left[Q_i^2- 2 u_i^2 + d_i^2 -
L_i^2+\ell_i^2\,\right].  
\eeq
The pure $U(1)_H^3$  and the mixed gravitational
anomalies can always be canceled by adding 
SM singlet fields with suitable charges,
and we assume they vanish. 
If the $C_n$'s in (\ref{coefficients}) 
do not vanish, the Green-Schwarz (GS) 
mechanism \cite{Green:1984sg} 
can be invoked to remove the anomalies 
by means of a $U(1)_H$ gauge shift 
of an axion  field $\eta(x)\to \eta(x) -\xi(x)\,\delta_{GS}$.  
The consistency conditions for this
cancellation read \cite{Ibanez:1993fy}  
\beq
\label{ibanez}
{C_1}/{k_1}={C_2}={C_3}=\delta_{GS},   
\eeq
where the Kac-Moody levels of the $SU(2)_L$ and 
$SU(3)_C$ gauge groups have been assumed to be unity 
and, since we are not postulating any GUT symmetry,   
the $U(1)_Y$ normalization factor $k_1$ is 
arbitrary. Then the weak mixing angle 
(at some large scale $\Lambda$) is  given by 
$\tan^2\theta_W=g^{\prime 2}/g^2=1/k_1\,$. 
Using  (\ref{coefficients}),  conditions  
(\ref{ibanez}) translate into 
\beq
\label{GScondition}
2(n_\phi-n_d+n_\ell) = (k_1 -\frac{5}{3})\> \delta_{GS}\,. 
\eeq
Now, one can assume that  
the gauge couplings unify for the canonical value 
$\tan^2\theta_W = {3}/{5}\,$ \cite{Nir:1995bu}. 
Then $n_\phi=n_d-n_\ell$ is obtained.
Alternatively, one can assume that 
for some reasons  the l.h.s. in (\ref{GScondition}) 
vanishes, and thus predict canonical gauge couplings 
unification \cite{Binetruy:1995ru}. 
However, in the absence of a GUT symmetry the value   
$k_1=5/3$ is not compelling. 
Other values of $k_1$ can be in reasonable agreement 
with unification at scales $\Lambda\neq \Lambda_{GUT}$ 
\cite{Ibanez:1993fy}, so that  $n_\phi$ and $n_d-n_\ell$  
are not necessarily related in any simple way. 
For a non-anomalous $U(1)_H$,  
(\ref{ibanez}) and (\ref{GScondition})  
still hold with $\delta_{GS}=0$, so that  
the interplay with 
gauge couplings unification is lost.   
However,   $n_\phi=n_d-n_\ell$ now follows as  an 
unavoidable consistency condition, giving a first 
constraint on the permitted horizontal charges.

Let us now study the symmetry properties of the 
coefficients  (\ref{coefficients}).
Since for each $SU(2)_L$ multiplet  
${\rm Tr}[T_3\,Y\,H]=Y\,H\,{\rm Tr}[T_3]=0$,    
the mixed electromagnetic-$U(1)_H$ anomaly 
can be expressed in terms of $C_1$ and $C_2$ as  
$C_Q = {1\over 2}(C_1+C_2)$.
Being $SU(3)_C\times U(1)_Q$  
vectorlike, it is free of $B$ and $L$ anomalies,
and then  $C_3$ and $C_Q$ must be invariant under shifts 
of the horizontal charges proportional to $B$ and $L$.   
Clearly, the same is not true for $C_1$ and $C_2$ 
separately. However, the SM is free of  $B$-$L$ anomalies,
and thus $C_1$ and $C_2$ are invariant 
under the corresponding shift.   
Also $L_\tau$-$L_\mu$ and $L_\mu$-$L_e$ have vanishing 
anomalies with $G_{SM}$, so they identify two more 
possible  shifts that leave invariant the $C_n$'s.   
In the following we state the consistency 
conditions for cancellation 
of the $G_{SM}\times U(1)_H$ gauge anomalies. 

A set of horizontal charges 
$\{H\}$ is equivalent to a second set $\{H''\}$ 
for which the coefficients $C_n''$ of   
the mixed linear anomalies vanish, if and only if 
the mixed $U(1)_Q^2$-$U(1)_H$ and $SU(3)_C^2$-$U(1)_H$ 
anomaly coefficients are equal:  
\begin{equation}
\label{condition1}
C_Q-C_3 = 0\qquad \Longleftrightarrow  \qquad  
C''_1=C''_2=C''_3=0. 
\end{equation}
Moreover, if for $\{H''\}$ the charge of the $\mu$ term 
$n''_\phi$ is different from zero, the 
coefficient of the quadratic anomaly 
$\widetilde C^{(2)}$ can always be set to zero: 
\beq
\label{condition2}
        n''_\phi \neq 0 
\qquad \Longrightarrow  \qquad  \widetilde C^{(2)} = 0.
\end{equation}
(As it stands, this condition is sufficient but not 
necessary. However, if all the neutrinos are mixed  
at a measurable level (\ref{condition2}) 
turns out to be necessary \cite{MNardiR}. 
In the following we take $n_\phi\neq 0$ 
in the strong sense). 

To prove this, let us assume that for the initial set
$\{H\}$ 
$C_n\neq0$. Then we start by shifting the charges  
proportionally  to the $X$ quantum numbers. 
$H\to H+{a\over 3}X$ yields:    
\beq
\label{shift1}
C_n \to  C'_n=C_n +\alpha_n\, a\,,
\eeq
with $\alpha_3=1$, $\alpha_2=-1/3$ and $\alpha_1=+7/3$. 
We  fix  $a=-C_3/\alpha_3$ so that $C'_3=0$.
Note that the combination 
$(C_1+C_2)/(\alpha_1+\alpha_2)-C_3/\alpha_3=C_Q-C_3$  
besides being $B$ and $L$ invariant, is  
also $X$ invariant by construction.  
Now a shift proportional to $B$ can be used to 
set $C''_2=0$. Since $C_3$ is $B$ 
invariant, $C''_3=C'_3=0$.
The sum $C'_1+C'_2$ is also  $B$ invariant and thus 
$C''_1=C'_1+C'_2=2\,C'_Q$. However,   
by assumption $C'_Q=C'_3\>(=0)$ and then the set $\{H''\}$ 
has vanishing mixed linear anomalies.  
Now, in order to cancel the quadratic anomaly
while keeping vanishing $C''_n$, 
we can use any of the  SM anomaly free 
symmetries   $B$-$L\,$,  $L_\tau$-$L_\mu\,,$  
$L_\mu$-$L_e$ (that in general will 
have non-vanishing mixed anomalies
with $U(1)_H$). Since  
$L_\tau$-$L_\mu$ and $L_\mu$-$L_e$ transform between 
non equivalent set of charges, we keep this freedom 
to account for two neutrino mixings 
(the third one results as a prediction) and     
we use $B$-$L$.
Under the charge redefinition $H \to H + \beta\,$($B$-$L$) 
\beqa
{C^{(2)}}''\to  \widetilde C^{(2)} &=&
{C^{(2)}}''+ \beta\, \left[
{4\over 3} n''_u - 
{2\over 3} n''_d + 
2 n''_\ell \right] \nnu \\
&=&  
{C^{(2)}}'' - 2\,\beta\,n''_\phi,  
\eeqa
where in the last step we have used the identity 
$ {4\over 3} n_u - {2\over 3} n_d + 
2 n_\ell =  C_1+C_2 - {4\over 3}\,C_3- 2\,n_\phi$
and the vanishing of the $C''_n$. 
If,  
as we have assumed, $n''_\phi\neq 0$,  
we can always set $\widetilde C^{(2)}=0$ by choosing   
$\beta=C^{(2)\prime\prime}/ (2\,n''_\phi)$.  
The constraint derived here is again stronger
than in the anomalous case: 
$C^{(2)}$ cannot be canceled with the GS 
mechanism, and one has to redefine the charges 
so that it vanishes identically. 
Assuming $k_1=\frac{5}{3}$, the GS consistency 
conditions (\ref{ibanez}) yield    
$C_1+C_2 - {4\over 3}\,C_3={4\over 3}C_3\neq0$ 
and then $C^{(2)}=0$ does not constrain 
the charges in any useful way. 

A set of horizontal charges $\{H\}$ for which 
$C_n=C^{(2)}=0$ identifies a one parameter 
family of anomaly free charges generated by 
shifts proportional to hypercharge: $H \to H+yY$. 
For the $C_n$'s this is trivial
due to the vanishing of the SM anomalies  
${\rm Tr}[SU(n)^2Y]=0$. For $C^{(2)}$ we have 
${\rm Tr}[Y\,H^2] \to {\rm Tr}[Y\,(H+Y)^2] = 2 C_1 =0 $. 

This property can be useful for model building:   
if the charges of the $i$-th family satisfy  
$L_i- d_i= u_i-Q_i=Q_i- \ell_i\,$, 
then it is possible to arrange the 
corresponding fermions into  
$\bf \bar 5$ + $\bf 10$ representations of $SU(5)$. 
Alternatively we can choose to fix   
e.g. $\phi_d=\phi_u=n_\phi/2$.  

In summary, imposing cancellation of the 
$G_{SM}\times U(1)_H$ gauge anomalies results in 
the following constraints on the fermions charges 
%
\begin{equation}
\label{mucharge}
n_\phi \neq 0,   \qquad 
n_\phi = n_d-n_\ell 
\simeq \ln_\lambda {\det M^d \over \det 
 M^\ell}\,,   
\end{equation}
%
where the last relation follows from 
(\ref{detmd}) and (\ref{detml}).
Since $n_d\neq n_\ell$ we conclude that 
{\it i) 
Yukawa coupling unification is permitted at most 
for two families}. 
Together with assumption 5,   
we also obtain  $n_\phi<0$ so that 
{\it ii) 
the  superpotential $\mu$ term is forbidden 
by holomorphy and vanishes in the supersymmetric limit}. 
Let us confront these results with phenomenology. 
To a good approximation the mass ratios 
$m_e/m_\mu \sim \lambda^{3\div 4}$, $m_\mu/m_\tau \sim \lambda^2$, 
$m_d/m_s \sim \lambda^2$ and $m_s/m_b \sim \lambda^2$
are renormalization group invariant. Then, since  
Yukawa coupling unification works remarkably well
for the third family, 
$\det M^\ell / \det M^d \sim \lambda$ or $\lambda^2$, and 
the allowed values of $n_\phi$  
are $-1$ or $-2\,$.  
Then a $\mu$ term arising from the (non-holomorphic) 
K\"ahler potential \cite{GMasiero} will have a value  
somewhat below the supersymmetry breaking scale $m_{3/2}$: 
\beq
\mu \sim \lambda^{|n_\phi|}\> m_{3/2}\qquad {\rm with} 
\qquad n_\phi=-1\>{\rm or}\>-2 \,.    
\eeq

As we have explicitly shown, the  anomaly cancellation 
condition $C_Q-C_3=0$  (\ref{condition1}) is $Y$, 
$B$, $L$ and $X$  invariant (as it should be), and hence 
it shares the same invariance of the Yukawa couplings.   
Therefore, any product of the determinants 
(\ref{detmu})-(\ref{detml}) for which the  
overall horizontal charge  can be recasted just 
in terms of the $C_n$'s must depend 
precisely on this  combination.
Such  a relation was first found in \cite{Nir:1995bu}.  
Given that $C_Q-C_3=n_\ell-{2\over 3}n_d+{1\over 3}n_u+n_\phi$
we can write it down at once:   
\beq
\label{combination}
\left({{\rm det} M^\ell\over \vev{\phi_d}^3}\right)
\left({{\rm det} M^d\over \vev{\phi_d}^3}\right)^{-{2\over3}}
\left({{\rm det} M^u\over \vev{\phi_u}^3}\right)^{1\over3}
\simeq \lambda^{C_Q-C_3}.
\eeq
Let us confront this relation with phenomenology. 
Anomaly cancellation implies that the r.h.s. 
is unity, while the  l.h.s. is bounded by an upper 
limit of order 
$\left[
\left({\rm det}M^d/ \vev{\phi_d}^3\right) 
\left({\rm det}M^u/\vev{\phi_u}^3\right)
\right]^{1/3}\ll 1$\cite{Nir:1995bu}. This inconsistency 
(or similar ones) led several authors to conclude 
that $U(1)_H$ must be anomalous 
\cite{Ibanez:1994ig,Binetruy:1995ru,GaugeU1,Nir:1995bu}.
However, (\ref{combination}) is meaningful 
only under the assumption that none of the determinants 
vanishes, and since low energy phenomenology is   
still compatible with a massless up quark  
\cite{Cohen:1999kk,mup0old} (see however 
\cite{mup0Leutwyler}) this might not be the case.
In the following we prove that 
insisting on the vanishing of the gauge anomalies 
yields $m_{\rm up}=0$ as a prediction.

We start by noticing that if the determinant 
of the matrix $U_{ij}\sim\lambda^{Q_i+ 
u_j+\phi_u}$ has an overall negative charge 
$ \eta^U  \equiv  n_u  +  3 \phi_u $ $\sim 
\log_\lambda\det U <0 $,  
then $M^u$ has vanishing eigenvalues. 
This is because $\det U$ consists of the sum 
of six terms of the form 
$\lambda^{n_1}\cdot\lambda^{n_2}\cdot\lambda^{n_3}$  
where $n_1+n_2+n_3= \eta^U<0$. Then 
at least one of the $n_i$ must be negative,    
corresponding to a holomorphic zero in the mass
matrix. Hence each one of the six terms vanishes.  

Now, if  $U(1)_H$ is anomaly free and assumption 5 holds,   
it is easy to see that 
{\it iii) the determinant of the six quark mass 
matrix ${\cal M}_q$ vanishes}: 
%
\beqa
\label{theorem2}
\rcasos{l} 
n_\ell \geq n_d  \    \\
C_n=C^{(2)}= 0 \   
\rfincasos
\quad \Longrightarrow  \quad  \det {\cal M}_q=0. 
\eeqa
%
In fact adding and subtracting $3n_\phi$ 
to $C_3=0$ yields   
\beq
 \eta^U +  \eta^D = 3\, n_\phi < 0. 
\eeq
Then at least one of the two  $\eta$  
must be negative, and the corresponding 
determinant vanishes. 
Of course, on phenomenological grounds, 
a massless up quark is the only 
viable possibility \cite{Cohen:1999kk,mup0old}. 
Using the $d$-quark mass ratios
given above, assuming  
$m_b/m_t\sim \lambda^3$ (as is preferred 
at large scales), 
and choosing $n_\phi=-1$, we obtain 
\beq
\label{etau}
 \eta^U   
\simeq   -9 - 3\,\log_\lambda\left(\frac{m_b}{m_t} \tan\beta\right),  
\eeq
that ranges between $-9$ and $-18$  for      
$\tan\beta$ between $m_t/m_b$ and 1. 
Because of the constraints from holomorphy,  
$ \eta^U<0$ results in an accidental $U(1)_u$ 
symmetry acting on the  $SU(2)_L$ singlet up quark:  
$u_1 \to e^{i\alpha} u_1$.  
Then the QCD $CP$ violating parameter 
$\bar\theta\equiv \theta +{\rm arg}\det{\cal M}_q$ 
is no more physical, and can be rotated away by means 
of a chiral transformation of the massless quark field. 
However, holomorphy is a crucial ingredient to 
achieve this result, and one has 
to check that  after supersymmetry is broken   
this result is not badly spoiled.
Supergravity effects induce  mixings in the 
kinetic terms. Canonical form is recovered 
by means of the field redefinitions  
$Q=V^Q Q'$ and $u=V^u u'$.  
Then the matrix of the Yukawa couplings  
$Y^u$ transforms into  ${Y^u}' = {V^Q}^T Y^u V^u$. 
Since $\det Y^u =0$,  $\det Y^{u\prime} =0$ follows,  
so that kinetic terms mixing, while it can lift  
mass matrix holomorphic zeroes, it does not lift 
vanishing eigenvalues.  
In general, soft supersymmetry breaking terms  
will not respect the $U(1)_u$ symmetry, 
so that  a mass for the up quark can  
be induced radiatively.
A conservative estimate of these effect gives  
$m_{\rm up} \lsim  (\alpha_s/\pi)\,
 \lambda^{|\eta^U-4|}\> \vev{\phi_u} 
\lsim 10^{-6}\,, (10)\,$eV [for $\tan\beta\sim 1\,, (m_t/m_b)$]  
where $\eta^U-4$ is the charge of the up-quark mass operator
when $m_c/m_t\sim \lambda^4\,$ is used.  
Following \cite{Pospelov:1999mv} we have estimated the possible   
contribution to the neutron electric dipole moment:  
$d_n/e\lsim 10^{-28}\,\bar\theta\>,(10^{-22}\,\bar\theta)\>$cm.  
Therefore, for moderate values of  $\tan\beta\,$, the neutron dipole
moment remains safely below the experimental limit 
$d_n/e < 6.3 \times 10^{-26}\>$cm \cite{Harris:1999jx} 
even for $\bar \theta\sim 1$.

Gauge symmetry and supersymmetry, together with 
constraints from fermion charges relations,  
imply that
{\it iv)
the superpotential has accidental 
$B$ and $L$ symmetries}.   
This result is deeply related  
to the solutions of the $\mu$ and strong CP 
problems ($n_\phi<0$, $\eta^U<0$).
The proof of {\it iv)} requires phenomenological 
inputs, like fermion mass ratios and  CKM mixings  
plus the assumption that neutrinos mixings are sizeable. 
Since it is somewhat lengthy, we   
will present it elsewhere \cite{MNardiR}.  
An intuitive (but not rigorous) argument 
goes as follows: 
given a set of minimal charges that fit well 
the fermion masses and mixings, the 
(shifts invariant) value of $\eta^U$ (\ref{etau})  
implies that $C^{(2)}$ in (\ref{quadratic}) 
is negative. To cancel $C^{(2)}$ the shift 
$H\to H+\beta\cdot$($B$-$L$) is required, 
where $\beta=C^{(2)}/2n_\phi$ is positive. 
All the R-parity violating operators 
$\mu_L L \phi_u$, 
$\lambda LL \ell\,$, 
$\lambda'LQ  d\,$ and  
$\lambda'' u d d\,$      
have $B$-$L$$=-1$, so that under this shift  
their charges are driven to negative values
implying that they are not allowed in the 
superpotential.  
Of course, dimension five see-saw operators for 
neutrino masses are also  forbidden. 
However, the same mechanism that generates 
$\mu$ will generate (with larger suppressions) 
also $\mu_L L\phi_u$ terms, which induce  
s-neutrinos VEVs. Canonical diagonalization of 
$L$-$\phi_d$ mixed kinetic terms will 
produce tiny $\lambda$ and $\lambda'$  
from the Yukawa couplings. Both these effects can result 
in small neutrino masses \cite{MNardiR}. However, 
since none of the $\lambda''$ can be generated 
in this way, proton stability is not in jeopardy.

Finally, let us stress that except for $\eta^U<0$     
the condition $C_Q-C_3=0$ does not imply other 
serious  constraints on charge assignments, so that   
a suitable choice of  horizontal charges can    
account for the observed pattern of fermion 
masses and mixings. 
The mass matrices of popular models \cite{Leurer:1993wg,GaugeU1} 
can be easily reproduced 
and, apart from $m_{\rm up}=0$, also the same 
phenomenology \cite{MNardiR}.


\medskip 

\centerline{\bf Acknowledgments.} 
\smallskip 
\noindent
We acknowledge conversations with J. Ferrandis 
and W. A. Ponce, and we thank Y. Nir for 
precious suggestions.
This work was supported by the DGICYT grant PB95-1077 
and by the TMR contract ERBFMRX-CT96-0090.
J. M. M. and D. A. R. are supported by Colciencias.



\end{document}